\documentclass[sigconf]{acmart}
\AtBeginDocument{%
  \providecommand\BibTeX{{%
    \normalfont B\kern-0.5em{\scshape i\kern-0.25em b}\kern-0.8em\TeX}}}

\setcopyright{acmcopyright}
\copyrightyear{2018}
\acmYear{2018}
\acmDOI{XXXXXXX.XXXXXXX}

\acmConference[Conference acronym 'XX]{Make sure to enter the correct
  conference title from your rights confirmation emai}{June 03--05,
  2018}{Woodstock, NY}
\usepackage{algorithm}
\usepackage{algorithmic}
\usepackage{booktabs}
\usepackage{multirow}
\usepackage{diagbox}
\newtheorem{definition}{Definition}
\usepackage{amsfonts}
\usepackage{amsmath}
\usepackage{balance}
\usepackage{subfig}
\usepackage{graphicx}

\begin{document}

\title{DynLLM: When Large Language Models Meet Dynamic Graph Recommendation}

\author{Ziwei Zhao}
\email{zzw22222@mail.ustc.edu.cn}
\affiliation{%
  \institution{University of Science and Technology of China}
  \city{Hefei}
  \country{China}
}

\author{Fake Lin}
\email{fklin@mail.ustc.edu.cn}
\affiliation{%
  \institution{University of Science and Technology of China}
  \city{Hefei}
  \country{China}
}

\author{Xi Zhu}
\email{xizhu@mail.ustc.edu.cn}
\affiliation{%
  \institution{University of Science and Technology of China}
  \city{Hefei}
  \country{China}
}

\author{Zhi Zheng}
\email{zhengzhi97@mail.ustc.edu.cn}
\affiliation{%
  \institution{University of Science and Technology of China}
  \city{Hefei}
  \country{China}
}

\author{Tong Xu}
\authornote{Corresponding Author.}
\email{tongxu@ustc.edu.cn}
\affiliation{%
  \institution{University of Science and Technology of China}
  \city{Hefei}
  \country{China}
}

\author{Shitian Shen}
\email{shitian.sst@taobao.com}
\affiliation{%
  \institution{Alibaba Group}
  \city{Hangzhou}
  \country{China}
}

\author{Xueying Li}
\email{xiaoming.lxy@taobao.com}
\affiliation{%
  \institution{Alibaba Group}
  \city{Hangzhou}
  \country{China}
}

\author{Zikai Yin}
\email{yinzikai@mail.ustc.edu.cn}
\affiliation{%
  \institution{University of Science and Technology of China}
  \city{Hefei}
  \country{China}
}

\author{Enhong Chen}
\email{cheneh@ustc.edu.cn}
\affiliation{%
  \institution{University of Science and Technology of China}
  \city{Hefei}
  \country{China}
}

\begin{abstract}
 Last year has witnessed the considerable interest of Large Language Models (LLMs) for their potential applications in recommender systems, which may mitigate the persistent issue of data sparsity. Though large efforts have been made for user-item graph augmentation with better graph-based recommendation performance, they may fail to deal with the dynamic graph recommendation task, which involves both structural and temporal graph dynamics with inherent complexity in processing time-evolving data. To bridge this gap, in this paper, we propose a novel framework, called DynLLM, to deal with the dynamic graph recommendation task with LLMs. Specifically, DynLLM harnesses the power of LLMs to generate multi-faceted user profiles based on the rich textual features of historical purchase records, including crowd segments, personal interests, preferred categories, and favored brands, which in turn supplement and enrich the underlying relationships between users and items. Along this line, to fuse the multi-faceted profiles with temporal graph embedding, we engage LLMs to derive corresponding profile embeddings, and further employ a distilled attention mechanism to refine the LLM-generated profile embeddings for alleviating noisy signals, while also assessing and adjusting the relevance of each distilled facet embedding for seamless integration with temporal graph embedding from continuous time dynamic graphs (CTDGs). Extensive experiments on two real e-commerce datasets have validated the superior improvements of DynLLM over a wide range of state-of-the-art baseline methods. 
  
\end{abstract}





\maketitle
\begin{sloppypar}
\section{Introduction}
Recommender systems have been recognized as a prototypical tool within online platforms, serving to mitigate the challenges of information overload and enabling users to navigate the overwhelming amount of data to discover interests, such as item recommendation in e-commerce platforms \cite{zhou2018deep, he2020lightgcn, wang2019neural}, friend recommendation in social network media \cite{ji2021you, wu2019neural, zhao2023time, zhu2023few}, and videos recommendation in mobile video applications \cite{gong2022real, zhao2019recommending, zhao2011integrating}. The primary goal of recommender systems is to accurately identify user preferences and predict potential interactions among a large number of candidate items. Collaborative filtering (CF) is the predominant paradigm in recommender systems and has been successfully integrated with various techniques \cite{koren2008factorization, he2017neural, li2016collaborative}, yielding significant advancements in the field. However, traditional CF methods are limited in their ability to capture high-order connectivity from user-item interactions, which has led to the development of graph-based recommendation methods \cite{wang2019neural, wu2021self, jin2020multi, zheng2021drug} inspired by Graph Convolutional Networks (GCNs) \cite{kipf2016semi}.

Unfortunately, the aforementioned graph-based recommendation methods are intrinsically constrained by their design to leverage static graph structures. As a consequence, these approaches exhibit a fundamental limitation: they are unable to account for and incorporate temporal information, which is crucial for understanding the evolving properties of user preferences and user-item interactions over time. To address this limitation, researchers have explored modeling temporal graphs as a sequence of discrete graph snapshots, defined as discrete time dynamic graphs (DTDGs), where each snapshot encapsulates the interactions that occur within a self-defined time interval \cite{pareja2020evolvegcn, sankar2020dysat}. Along this line, a nontrivial challenge in DTDGs is how to choose an optimal time interval especially when encountering large-scale dynamic graphs. In pursuit of a more comprehensive integration of fine-grained temporal information and topological graph structure, the paradigm of continuous time dynamic graphs has been introduced. Specifically, temporal point processes (TPPs) which consider the event time as a random variable allow for the precise modeling of temporal information at a fine-grained level \cite{zuo2018embedding, lu2019temporal}. Subsequent advancements in this field, as exemplified by the work of Xu et al. \cite{xu2020inductive} and its further extension by Rossi et al. \cite{tgn_icml_grl2020}, have introduced functional time encoding technique and temporal attention mechanism to refine the representation of fine-grained temporal information and aggregate messages across neighboring nodes within the temporal graph context, which has achieved notable improvements in CTDGs. 

\begin{figure}[t]
\centering
\includegraphics[width=0.95\columnwidth]{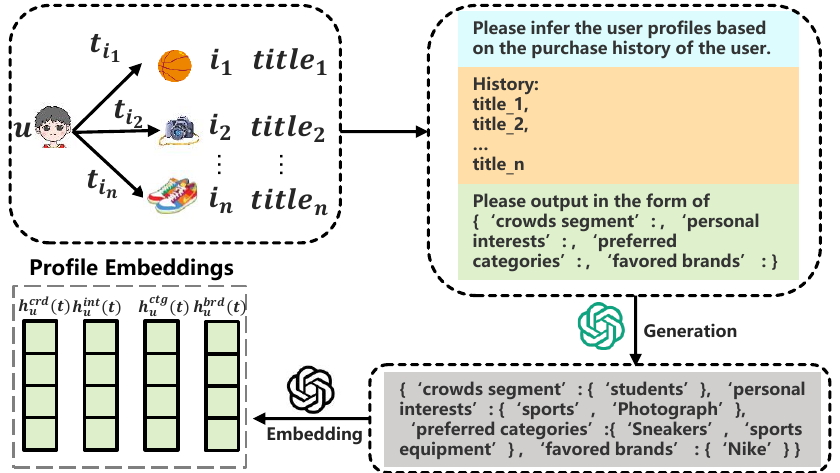} 
\caption{The process of generating profile embeddings from LLMs. According to the rich textual titles $\{title_1, title_2, \dots, title_n\}$ from corresponding purchased items $\{i_1, i_2, \dots, i_n\}$ of a user $u$, the multi-faceted profiles are derived from LLM-generation through designed prompt, and then the profile embeddings are generated through LLM-embedding.}
\label{motivation}
\end{figure}

Despite these improvements, challenges such as data sparsity and limited interpretability persist in both static and dynamic graph recommendations. The rise of Large Language Models (LLMs), typified by AI generators like ChatGPT, has reshaped human-computer interaction and shown promise in recommender systems. Specifically, LLMs enhance recommendation by providing rich language generation capabilities that contribute to solving data sparsity \cite{hou2023large, wang2023zero} and improve interpretability \cite{gao2023chat, zheng2023generative}. Regrettably, LLMs may not inherently possess the ability to understand graph data directly for user-item recommendation. Recent efforts have been directed towards augmenting graph-based recommendation through the generation of side information \cite{wei2023llmrec} and the construction of meta-path prompts \cite{wu2023exploring}, which aim to address data sparsity and uncover underlying patterns in graph recommendation scenarios.

However, the integration of LLMs within dynamic recommendation has remained unexplored, primarily due to the complexity of adapting LLMs to forecast dynamically changing data \cite{sun2023test}. In this work, we pioneer the seamless fusion of LLMs with dynamic recommendation in CTDGs. 
Specifically, We initiate the DynLLM framework by employing temporal graph attention networks (TGANs) to aggregate embeddings from temporal neighbors for users and items, which efficiently captures both fine-grained temporal dynamics and intricate graph topology information. Subsequently, as depicted in Figure \ref{motivation}, we utilize LLMs to generate comprehensive user profiles from textual features in purchased item titles, which are segmented into four distinct components: crowd segments, personal interests, preferred categories, and favored brands. Through the capability of semantic representation generation, LLMs are also engaged to derive corresponding profile embeddings. As indicated in \cite{wei2023llmrec}, augmented data from LLMs may be contaminated with noise, which potentially leads to making optimization unstable and ineffective. In our case, to mitigate the influence of generated noise and emphasize relevant signals, we present an innovative distilled attention mechanism to refine and integrate these multi-faceted profile embeddings. This mechanism first distills the profile embeddings generated from LLMs, efficiently minimizing noise and enhancing expressiveness. It then employs a multi-head attention to merge the distilled profile embeddings with the temporal graph embedding from CTDGs, thereby illustrating which profile facet contributes more to modeling the evolving user preferences. Finally, user and item embedding are updated for next-time recommendation through the iterative nature of DynLLM, integrating not only graph topology information but also fine-grained temporal information and the generated rich textual information. 

Overall, our major contributions are summarized as follows:
\begin{itemize}
\item To the best of our knowledge, we are the first to integrate LLMs with dynamic recommendation, utilizing the framework of CTDGs. This integration offers a novel perspective on dynamically modeling user preferences and user-item interactions.
\item We introduce a strategy based on LLMs for multi-faceted information augmentation, aiming to capture a comprehensive view of user profiles.
\item We design an innovative multi-faceted distilled attention mechanism. This mechanism serves a dual purpose: it refines the profile embeddings by mitigating the noise generated by LLMs, and it seamlessly incorporates the refined multi-faceted profile embeddings with temporal graph embedding by weighting their contribution to the evolving users representation.
\item We conduct rigorous experiments on two real-world datasets collected from online e-commerce platforms, which significantly proves that the proposed DynLLM surpasses state-of-the-art models in dynamic recommendation scenarios.
\end{itemize}

\section{Related Work}
In this section, we review the literature pertinent to our work, categorizing it into three parts: static graph recommendation, dynamic graph recommendation, and LLM-based recommendation.
\subsection{Static Graph Recommendation}
The past decade has witnessed a surge in research on graph embedding techniques, driven by the prevalence of network structures across various domains. Early works such as LINE \cite{tang2015line} and node2vec \cite{grover2016node2vec} leverage random walk strategies to capture node similarity and learn node representations effectively. Advancements in deep learning have subsequently extended to graph data, with the introduction of Graph Convolutional Networks (GCNs) \cite{kipf2016semi}, NGCF \cite{wang2019neural} adapts GCNs to the user-item interaction graph to capture CF signals by deepening the use of subgraph structure with high-hop neighbors, and LightGCN \cite{he2020lightgcn} only include the most essential neighborhood aggregation component in GCNs for CF, where the feature transformation and nonlinear activation contribute little to the performance of CF are discarded. With the emergence of advanced techniques, researchers have endowed GNN-based recommendation with meta-path learning \cite{wang2019heterogeneous, fan2019metapath}, hypergraph learning \cite{jia2021hypergraph, la2022music} and contrastive learning \cite{jiang2023adaptive, chen2023heterogeneous}, leading to remarkable progress in the field of static graph recommendation domain.

\subsection{Dynamic Graph Recommendation}
Dynamic graph recommendation builds upon the foundations of static graph recommendation techniques by integrating additional temporal information. In terms of time granularity, dynamic graphs can be categorized as discrete time dynamic graphs (DTDGs) and continuous time dynamic graphs (CTDGs). For recommendation in DTDGs, DySAT \cite{sankar2020dysat} computes dynamic node representations through joint self-attention along two dimensions of structural neighborhood and temporal dynamics. To address the challenge of node dynamics, such as node addition and deletion, EvolveGCN \cite{pareja2020evolvegcn} innovatively allows the evolution of GCN parameters over time without resorting to node embeddings. Nevertheless, how to determine an optimal time granularity for DTDGs remains an open question. In the realm of CTDGs, deepcoevolve \cite{dai2016deep} and $\textrm{M}^2\textrm{DNE}$ \cite{lu2019temporal} both employ temporal point processes (TPPs) to model the interaction formation of user-item graph, JODIE \cite{kumar2019predicting} addresses the challenge of embedding staleness by introducing a linear embedding projector within a coupled RNN framework. To enhance the integration of fine-grained temporal information, TGAT \cite{xu2020inductive} introduces a sophisticated time encoding module from Bochner's theorem in harmonic analysis. Furthermore, TGN \cite{tgn_icml_grl2020} proposes an innovative memory module for storing past user-item graph information and forecasting future interactions.

\subsection{LLM-based Recommendation}
Large Language Models (LLMs) have gained prominence in the domain of recommender systems \cite{wu2023survey}, catalyzing significant progress in the field. So far, Wang et al. \cite{wang2022towards} proposes a unified conversational recommender system that leverages knowledge-enhanced prompt learning grounded in pre-trained language models (PLMs). In the context of challenges such as cold-start problems and cross-domain recommendation scenarios, Chat-REC \cite{gao2023chat} makes the recommendation process more interactive and explainable through in-context learning between users and items. Addressing the issue of sparse recommendation data in LLMs' pre-training period, TALLRec \cite{bao2023tallrec} emerges as a solution that fine-tunes LLMs for downstream recommendation tasks, which ensures that LLMs can still generate high-quality recommendation despite limited pre-training data. Furthermore, the integration of LLMs has begun to show promising results in graph-based recommendation. LLMRec \cite{wei2023llmrec} enhances graph recommendation by introducing three efficient LLM-based graph augmentation strategies, and Wu et al. \cite{wu2023exploring} proposes a meta-path prompt constructor to utilize LLMs to interpret complex behavior graphs.

\section{Preliminaries}
In this section, we first introduce graph recommendation and continuous time dynamic graphs (CTDGs), and then provide the definition of dynamic recommendation with CTDGs.

\begin{table}[t]
\centering
\caption{Basic mathematical notations}
\begin{tabular}{cl}
\toprule
Symbol & Description \\ 
\midrule
$\mathcal{T}$, $t$ & The set of timestamps and $t \in \mathcal{T}$  \\ 
$U_t$ & The set of users at $t$ \\ 
$I_t$ & The set of items at $t$ \\ 
$Y_t$ & The interaction matrix at $t$ \\
$(u, i, t)$ & An interaction of user $u$ and item $i$ at $t$ \\
$title_i$ & The textual title of item $i$ \\
$\boldsymbol{h}_{u}(t)$, $\boldsymbol{h}_{i}(t)$ & The embedding of $u$ and $i$ at $t$ \\ 
$\boldsymbol{h}_u^{\mathcal{N}}(t)$, $\boldsymbol{h}_i^{\mathcal{N}}(t)$ & The neighboring embedding of  $u$ and $i$ at $t$ \\
$\boldsymbol{h}_u^G(t)$, $\boldsymbol{h}_i^G(t)$ & The temporal graph embedding of $u$ and $i$ at $t$ \\
$\boldsymbol{h}_u^A(t)$ & The distilled profile embedding of $u$ at $t$  \\
$\boldsymbol{\Bar{h}}_i$ & The static embedding of item $i$ \\ 
$\boldsymbol{\Tilde{h}}_i(t)$ & The dynamic embedding of item $i$ at $t$ \\ 
\bottomrule
\end{tabular}

\label{notations}
\end{table}

\subsection{Graph Recommendation} 
\begin{definition}
    \textbf{Graph Recommendation} Given a graph  $G = (U, I, Y)$, $U = \{u_1, u_2, \dots, u_M\}$ denotes the set of users and $I = \{i_1, i_2, \dots, i_N\}$ denotes the set of items, in which $M$ and $N$ are the number of users and items, respectively. $Y \in \{0, 1\}^{M \times N}$ is the interaction matrix, if there is an interaction between user $u$ and item $i$, then $Y_{ui} = 1$, otherwise $Y_{ui} = 0$. Then graph recommendation is to learn a scoring function $g$ that estimates the likelihood that a user $u$ will interact with an item $i$, i.e., $g_{(u, i)}: (u, i) \rightarrow s$, where $s \in \mathbb{R}$.
\end{definition}
\subsection{Continuous Time Dynamic Graph} 
\begin{definition}
    \textbf{Continuous Time Dynamic Graphs (CTDGs)} Given $G = \left( U_t, I_t, Y_t, \mathcal{T}\right)$, $U_t$ and $I_t$ are separately the users and items at timestamp $t \in \mathcal{T}$, $\mathcal{T}$ denotes the timestamp sequence set belonging to $\mathbb{R}^+$, and $Y_t \subseteq U_t \times I_t \times \mathcal{T}$ denotes the temporal interactions formed at timestamp $t$. By this definition, each interaction  $(u_k, i_k, t)$ can be assigned with a unique timestamp in the finest granularity.
\end{definition}

\subsection{Dynamic Recommendation with CTDGs}
We first list some important mathematical notations in Table \ref{notations} used throughout this paper. Accordingly, the problem of dynamic recommendation with CTDGs can be formulated as follows:
\begin{definition}
\textbf{Dynamic Recommendation with CTDGs.} Given a continuous time dynamic graph $G = \left( U_t, I_t, Y_t, \mathcal{T}\right)$ as formulated before, the goal of dynamic recommendation with CTDGs is to acquire a personalized score function to predict the likelihood if an interaction $(u, i, t)$ will be formed at a specific timestamp $t$. Formally speaking, dynamic recommendation aims to learn a function $f$, s.t. $f_{(u, i, t)}: (u, i, t) \rightarrow s$, where $s \in \mathbb{R}$. And the candidate items are ranked in descending order according to the score $s$ to generate the Top-K item recommendation list.
\end{definition}

\begin{figure*}[htbp]
\centering
\includegraphics[width=1.85\columnwidth]{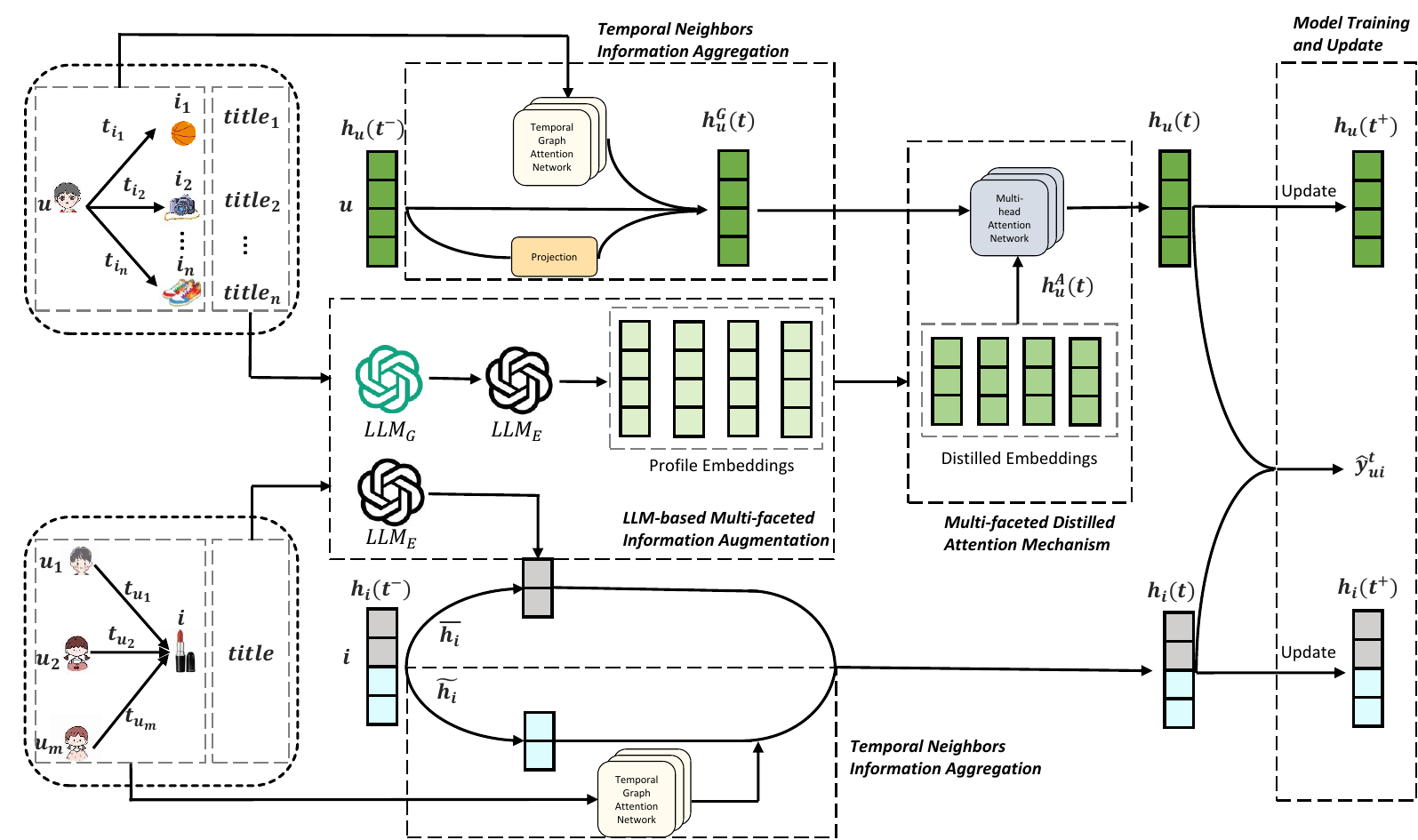} 
\caption{The overall framework of the proposed DynLLM. There are four principal modules: (a) Temporal Neighbors Information Aggregation that aggregates temporal neighbors information for users and items in CTDGS, (b) LLM-based Multi-faceted Information Augmentation that generates multi-faceted user profiles through LLMs, (c) Multi-faceted Distilled Attention Mechanism that distills and incorporate LLM-generated embeddings to enhance representation, and (d) Model Training and Update that trains the model and update embeddings for future recommendation.}
\label{fig3}
\end{figure*}

\section{Methods}
In this section, we will illustrate the intricacies of our proposed framework, designated as \textbf{DynLLM}, which harnesses LLMs to enhance dynamic graph recommendation. Our model is architecturally composed of four principal modules, i.e., (a) Temporal Neighbors Information Aggregation, (b) LLM-based Multi-faceted Profiles Augmentation,  (c) Multi-faceted Distilled Attention Mechanism, and (d) Model Training and Update. Initially, we employ temporal graph attention networks (TGANs) to meticulously aggregate temporal neighbors information for both users and items in CTDGs. Then we leverage LLMs to construct multi-faceted user profiles through textual titles of historical purchase items, guided by carefully designed prompts. These profiles include various dimensions such as crowd segments, personal interests, preferred categories, and favored brands, thereby enriching the user representation with a multi-faceted perspective. To further refine this representation, we introduce a novel multi-faceted distilled attention mechanism. This mechanism is responsible for the distillation of multi-faceted profile embeddings derived from LLMs, ensuring the reduction of noise and the enhancement of signal relevance. In parallel, it also integrates the distilled embeddings with temporal graph embedding from CTDGs, emphasizing the importance of each profile facet.
Lastly, the user embedding and item embedding are mixed for prediction and then updated through Gated Recurrent Unit (GRU) updaters, preparing for next-time recommendation.

\subsection{Temporal Neighbors Information Aggregation}
In this module, to aggregate temporal neighbors information of both users and items, TGANs based on CTDGs are utilized to generate neighboring embedding $h_u^{\mathcal{N}}$ and $h_i^{\mathcal{N}}$, respectively. For simplicity, we only show how to derive $\boldsymbol{h}_u^{\mathcal{N}}$. First, we define the temporal neighbors of a specific user $u$ at timestamp $t$ as $\mathcal{N}_u(t)$ as follows:
\begin{align}
    \mathcal{N}_u(t) &= \{(i_1, t_1), (i_2, t_2), \cdots, (i_n, t_n)\}. 
\end{align}
Here $(i_1, t_1), (i_2, t_2), \cdots, (i_n, t_n)$ means user $u$ interacted with items $i_1, i_2, \cdots, i_n$ at timestamps $t_1, t_2, \cdots, t_n$, respectively.

Subsequently, given $\{\boldsymbol{h}_{i_1}^{l-1}(t^-), \boldsymbol{h}_{i_2}^{(l-1)}(t^-), \cdots, \boldsymbol{h}_{i_n}^{(l-1)}(t^-)\}$ as embeddings of neighbors exactly before timestamp $t$ (denoted as $t^-$) at previous layer $l-1$, TGANs are employed to incorporate these embeddings of neighbors to generate $\boldsymbol{h}_u^{\mathcal{N}}(t)$ at timestamp $t$, the formula is listed as follows:
\begin{equation}
\label{temporal layer}
\begin{aligned}
    \boldsymbol{e}_k^{(l-1)}  & = \boldsymbol{h}_{i_k}^{(l - 1)}(t^-) \mathbin\Vert \phi (t - t_{k}), \\
    \boldsymbol{K}^{(l)}(t) & = \boldsymbol{V}^{(l)}(t) = \boldsymbol{e}_1^{(l-1)} \mathbin\Vert \boldsymbol{e}_2^{(l-1)} \mathbin\Vert \cdots \mathbin\Vert \boldsymbol{e}_n^{(l-1)} , \\
    \boldsymbol{q}^{(l)}(t) & = \boldsymbol{h}_u^{\mathcal{N}, (l - 1)}(t) \mathbin\Vert \phi(0),  \\
    \boldsymbol{h}_u^{\mathcal{N}, (l)} (t) & = \text{Attn}^{(l)}\left(\boldsymbol{q}^{(l)} (t), \boldsymbol{K}^{(l)}(t), \boldsymbol{V}^{(l)}(t)\right).
\end{aligned}
\end{equation}
Here, $\boldsymbol{h}_u^{\mathcal{N}, (0)}(t) = \boldsymbol{h}_u(t^-) + \boldsymbol{f_u}$ and $\boldsymbol{h}_i^{(0)}(t) = \boldsymbol{h}_i(t^-) + \boldsymbol{f_i}$, where $\boldsymbol{f}_u$ and $\boldsymbol{f}_i$ are the node features of users and items. And it is worth noting that the time encoding function $\phi(t)$ in the temporal edge embedding $\boldsymbol{e}_k^{(l-1)}$ is borrowed from \cite{xu2020inductive}:
\begin{align}
    \phi(t) = \sqrt{\frac{1}{d}}[\cos(\omega_1t), \sin(\omega_1t), \dots, \cos(\omega_dt), \sin(\omega_dt)], 
\end{align}
where $\omega_1, \omega_2, \cdots, \omega_d$ are the learnable frequency parameters and $d$ is the mapping dimension.

After stacking $L$ layers of TGANs, the neighboring embedding of user $u$ is computed as $\boldsymbol{h}_u^{\mathcal{N}}(t) = \boldsymbol{h}_u^{\mathcal{N}, (L)}(t)$, i.e., the last layer output of TGANs. Similarly, the neighboring embedding of item $i$ is $\boldsymbol{h}_i^{\mathcal{N}}(t) = \boldsymbol{h}_i^{\mathcal{N}, (L)}(t)$. 

Then, we borrow the aggregation method from \cite{zhao2023time} to acquire $\boldsymbol{h}_u^G(t)$ and $\boldsymbol{h}_i^G(t)$, which contains temporal information and graph topology information in CTDGs, denoted as temporal graph embedding of $u$ and $i$. The computation is executed as follows:
\begin{align}
    \boldsymbol{h}_u^G(t) &=  \textrm{MLP}  \left(  \boldsymbol{h}_u(t^-) \mathbin\Vert \boldsymbol{h}_u^{\mathcal{N}}(t) \mathbin\Vert  \boldsymbol{h}_u^{p}(t)\right), \\
\label{dynamic item}
    \boldsymbol{{h}_i^G}(t) & =  \textrm{MLP}  \left(  \boldsymbol{\Tilde{h}}_i(t^-) \mathbin\Vert \boldsymbol{h}_i^{\mathcal{N}}(t) \right).
\end{align}
Here $\boldsymbol{h}_u^p(t)$ is the projection embedding to avoid embedding staleness from \cite{zhao2023time}, $\boldsymbol{\Tilde{h}}_i(t^-)$ is the item dynamic embedding at $t^-$.

\subsection{LLM-based Multi-faceted Information Augmentation}
With leveraging the knowledge base and reasoning abilities of LLMs, we introduce a strategy that utilizes LLMs for multi-faceted information augmentation, which encapsulates user profiles from diverse perspectives by utilizing purchased item titles. 
And the augmented profiles also provide additional information to help recommend future items for users in turn. To effectively bridge the gap between LLMs and dynamic graph recommendation, our approach involves the dynamic generation of multi-faceted user profiles according to the changing history of purchased items.

Specifically, without loss of generality, suppose the purchased items history of user $u$ from timestamp $t - \Delta t$ to timestamp $t$ are $\{i_1^u, i_2^u, \cdots, i_n^u\}$, and the corresponding item titles are $\{title_1^u, title_2^u, \cdots, title_n^u\}$. Furthermore, to facilitate the generation of high-quality user profiles, a multi-faceted prompt $\mathcal{P}_u^{t- \Delta t, t}$ is constructed for LLM's completion, and the illustration is provided in Figure \ref{motivation}. Formally speaking, the procedure can be described as:
\begin{equation}
\label{prompt}
\begin{aligned}
    \textrm{crd}_u^{t - \Delta t, t}, \textrm{int}_u^{t - \Delta t, t}, & \textrm{ctg}_u^{t - \Delta t, t}, \textrm{brd}_u^{t - \Delta t, t} = \textrm{LLM}_G(\mathcal{P}_u^{t- \Delta t, t}),
\end{aligned}
\end{equation}
where $\textrm{crd}_u^{t - \Delta t, t}, \textrm{int}_u^{t - \Delta t, t},  \textrm{ctg}_u^{t - \Delta t, t}, \textrm{brd}_u^{t - \Delta t, t}$ are separately the crowd segments, personal interests, preferred categories and favored brands generated by $\textrm{LLM}_G$ (i.e., the LLM-generation model), which reflects the user profiles of $u$ from $t - \Delta t$ to $t$. 

Leveraging the proficiency of LLMs in text encoding, we proceed to input the augmented profiles into LLMs to obtain multi-faceted profile embeddings. Afterward, we employ fully connected MLPs with dropout to transform the multi-faceted embeddings, which not only reduces dimension but also projects the embeddings to their own space, the mathematical formulation is as follows:
\begin{equation}
\label{prompt embedding}
\begin{aligned}
    \boldsymbol{h}_{u}^{ \textrm{crd}}(t) & = \textrm{MLP}\left(\textrm{LLM}_E(\textrm{crd}_u^{t - \Delta t, t})\right), \\
    \boldsymbol{h}_{u}^{ \textrm{int}}(t) & = \textrm{MLP}\left(\textrm{LLM}_E(\textrm{int}_u^{t - \Delta t, t})\right), \\
    \boldsymbol{h}_{u}^{ \textrm{ctg}}(t) & = \textrm{MLP}\left(\textrm{LLM}_E(\textrm{ctg}_u^{t - \Delta t, t})\right), \\
    \boldsymbol{h}_{u}^{ \textrm{brd}}(t) & = \textrm{MLP}\left(\textrm{LLM}_E(\textrm{brd}_u^{t - \Delta t, t})\right).
\end{aligned}
\end{equation}
Here $\boldsymbol{h}_{u}^{ \textrm{crd}}(t), \boldsymbol{h}_{u}^{ \textrm{int}}(t), \boldsymbol{h}_{u}^{ \textrm{ctg}}(t)$ and $\boldsymbol{h}_{u}^{ \textrm{brd}}(t)$ are the transformed profile embeddings, and $\textrm{LLM}_E$ is LLM-embedding model.

As for items, with attributed titles $\textrm
{title}_i$ for item $i$, we feed it into LLMs and a subsequent MLP layer to acquire the item static embedding:
\begin{align}
\label{item llm}
    \boldsymbol{\Bar{h}}_i &= \textrm{MLP}\left(\textrm{LLM}_E(title_i)\right).
\end{align}

\subsection{Multi-faceted Distilled Attention Mechanism}
Upon obtaining the transformed multi-faceted profile embeddings, how to effectively incorporate these embeddings into dynamic user embedding is thought-provoking. To enhance the representation of transformed multi-faceted embeddings and reduce noisy signals generated by LLMs, we first distill these embeddings to mitigate the impact of generated noise. Formally, we take $\boldsymbol{h}_{u}^{ \textrm{crd}}(t)$ as example, and we define a parameter vector $\boldsymbol{q}$ to compute the significance weights for each dimension within $\boldsymbol{h}_{u}^{ \textrm{crd}}(t)$. Subsequently, dimensions corresponding to the top $r$ weights are selected and are distilled for output, and we denote $r$ as the distillation coefficient. The formal expression is:
\begin{equation}
\label{selection}
\begin{aligned}
    \boldsymbol{s} & = \boldsymbol{h}_{u}^{ \textrm{crd}}(t) \boldsymbol{q} / \Vert \boldsymbol{q} \Vert \\
    j & = \textrm{top-indices}(\boldsymbol{s}, r) \\
    \boldsymbol{\Tilde{h}}_{u}^{ \textrm{crd}}(t) & = [\boldsymbol{h}_{u}^{ \textrm{crd}}(t) \circ \textrm{tanh}(\boldsymbol{s})]_{j},
\end{aligned}
\end{equation}
where $\boldsymbol{\Tilde{h}}_{u}^{ \textrm{crd}}(t)$ is the distilled crowd embedding for user $u$ and $\textrm{tanh}$ is the activation function. Along this line, the distilled interests embedding $\boldsymbol{\Tilde{h}}_{u}^{ \textrm{int}}(t)$, the distilled categories embedding $\boldsymbol{\Tilde{h}}_{u}^{ \textrm{ctg}}(t)$ and the distilled brands embedding $\boldsymbol{\Tilde{h}}_{u}^{ \textrm{brd}}(t)$ can be computed in a similar way. 

Following this, it is essential to incorporate the LLM-generated information with the graph-based and temporal information. To show the contribution of each user profile facet, we commence by concatenating embeddings of each facet to form an overall LLM-augmented profile embedding $\boldsymbol{h}_u^{A}(t)$, then apply a multi-head attention to weigh the relevance of $\boldsymbol{h}_u^{A}(t)$ and $\boldsymbol{h}_u^{G}(t)$, and the specific calculation is:
\begin{equation}
\label{multi-head attention}
\begin{aligned}
    \boldsymbol{\Tilde{q}}^{(l)}(t) & = \boldsymbol{h}_u^{G}(t), \\
    \boldsymbol{h}_u^{A}(t) & = \boldsymbol{\Tilde{h}}_{u}^{ \textrm{crd}}(t) \mathbin\Vert \boldsymbol{\Tilde{h}}_{u}^{ \textrm{int}}(t) \mathbin\Vert \boldsymbol{\Tilde{h}}_{u}^{ \textrm{ctg}}(t) \mathbin\Vert \boldsymbol{\Tilde{h}}_{u}^{ \textrm{brd}}(t), \\
    \boldsymbol{\Tilde{K}}^{(l)}(t) & = \boldsymbol{\Tilde{V}}^{(l)}(t) = \boldsymbol{h}_u^{A}(t), \\
    \boldsymbol{h}_u^{(l)}(t) &= \textrm{Attn}^{(l)}(\boldsymbol{\Tilde{q}}^{(l)}(t), \boldsymbol{\Tilde{K}}^{(l)}(t), \boldsymbol{\Tilde{V}}^{(l)}(t)), 
\end{aligned}
\end{equation}
in which $\boldsymbol{h}_u(t) = \boldsymbol{h}_u^{(L)}(t)$ after computing $L$ layers multi-head attention. As a result, $\boldsymbol{h}_u(t)$ is a fusion of graph topology information, fine-grained time information, and LLM-based augmented information.
And the embedding of $i$ at timestamp $t$ is $\boldsymbol{h}_i(t) = \boldsymbol{\Bar{h}}_i \mathbin\Vert \boldsymbol{h}_i^G(t)$, here $\boldsymbol{h}_i^G(t)$ from Equation\ref{dynamic item} is also the item dynamic embedding $\boldsymbol{\Tilde{h}}_i(t)$ at timestamp $t$.


\subsection{Model Training and Update}
Through the above modules, we have gradually obtained conclusive representations of users and items at a specific timestamp, and the preference score for a user $u$ toward an item $i$ at the timestamp $t$ is defined as the inner product of each representation:
\begin{align}
    \hat{y}_{ui}^t &= \boldsymbol{h}_u(t)^\top \boldsymbol{h}_i(t),
\end{align}
which is used as the ranking score for recommend generation. Then followed by classical methods, we employ Bayesian Personalized Ranking (BPR) loss to train the whole model, which assumes that observed interactions will be assigned with higher scores than unobserved counterparts:
\begin{align}
    \ell = \sum_{(u, i^+, i^-, t) \in E} -\log \sigma (\hat{y}_{ui^+}^t - \hat{y}_{ui^-}^t) + \lambda \Vert \Theta \Vert^2.
\end{align}
Here $E$ is the set of training interactions, $i^+$ and $i^-$ are separately positive and negative samples from training data, weight decay regularization $\Vert \Theta \Vert ^2$ weighted by the coefficient $\lambda$ is to prevent overfitting, and $\sigma(\cdot)$ is activation function sigmoid to introduce non-linearity.

In order to perform next-time recommendation, we utilize GRUs to integrate current-time interaction information to update user and item embedding, which can be calculated as:
\begin{align}
    \boldsymbol{h}_u(t^+) & = \textrm{GRU}(\boldsymbol{h}_u(t), \boldsymbol{h}_i(t) \mathbin\Vert \phi(t - t_u^\prime)) \\
    \boldsymbol{\Tilde{h}}_i(t^+) & = \textrm{GRU}(\boldsymbol{h}_i^G(t), \boldsymbol{h}_u(t) \mathbin\Vert \phi(t - t_i^\prime)) \\
    \boldsymbol{h}_i(t^+) & = \boldsymbol{\Bar{h}}_i \mathbin\Vert \boldsymbol{\Tilde{h}_i}(t^+).
\end{align}
Here $t_u^\prime$ is the last timestamp at which $u$ interacted with any item, whereas $t_i^\prime$ denotes the last timestamp at which $i$ was interacted with by any user. Eventually, the updated embeddings $\boldsymbol{h}_u(t^+)$ and $\boldsymbol{h}_i(t^+)$ would be employed for upcoming next-time recommendation.

\section{Experiments}
In this section, we conduct comprehensive experiments on a real-world dataset from an online e-commerce platform to demonstrate the effectiveness of our proposed DynLLM framework. Specifically, we address the following research questions:
\begin{itemize}
    \item \textbf{RQ1} How does DynLLM compare to state-of-the-art methods, including static and dynamic graph recommendation models?
    \item \textbf{RQ2} How do the individual components of DynLLM, such as LLM-based multi-faceted information augmentation and multi-faceted distilled attention mechanism, enhance dynamic recommendation performance?
    \item \textbf{RQ3} What is the contribution of each facet of LLM-augmented profiles to the overall performance?
    \item \textbf{RQ4} What impact do the number of user neighbor ($\mathcal{N}_u$) and item neighbors ($\mathcal{N}_i$) have on the performance of DynLLM?
    \item \textbf{RQ5} How sensitive is the model to different distillation coefficient $r$?
\end{itemize} 

\subsection{Experimental Settings}
In the following parts, we will introduce our experimental settings, which include data description, baselines, reproductive settings, and evaluation protocols. 

\subsubsection{Data Description} 
We collected the purchase data from two online e-commerce platforms, covering the period from November 14, 2021, to November 14, 2023.
Initially, we sampled a fixed number of users and their corresponding interactions with items to create a subset of the dataset. Subsequently, we retained users who made purchases of over 20 distinct items, and items that were purchased by over 5 distinct users. This process was repeated until the number of users and items converged. It is important to note that each interaction was assigned a timestamp to demonstrate its temporal dynamics. Further details of the collected datasets can be found in Table \ref{stat}.

\begin{table}[htbp]
\centering
\caption{The statistics of the dataset.}
\label{stat}
\begin{tabular}{c|c|c}
\hline
Dataset     & Tmall & Alibaba \\ 
\hline
Number of users   &   5651 & 16901 \\ 
Number of items    &   10333  & 30080\\
Number of interactions & 258390 & 722716 \\
The average length of a title & 35.51 & 34.22 \\
The average interactions of a user & 45.72 & 42.76 \\
The average interaction of an item & 25.01 & 24.03 \\

\hline
\end{tabular}
\end{table}

\begin{table*}[htbp]
  \centering
  \caption{The overall performance is evaluated in terms of Recall@K and NDCG@K (abbreviated as R@K and N@K). Bond numbers indicate the best performance, while underlined numbers represent the best baseline performance. The \textit{\%Imp} denotes the improvements achieved by DynLLM relative to the runner-up.}
  \label{overall performance}
  \setlength{\tabcolsep}{2mm}{
    \begin{tabular}{c|cccccc|cccccc}
    \toprule
    {Dataset}                       & \multicolumn{6}{c|}{Tmall}     & \multicolumn{6}{c}{Alibaba}     \\ 
    \midrule
    Model    & R@10 & R@20 & R@30 & N@10 & N@20 & N@30 & R@10 & R@20 & R@30 & N@10 & N@20 & N@30 \\
    \midrule
    GCN     & 0.0287    & 0.0381    & 0.0610    & 0.0218    & 0.0241 & 0.0290 & 0.0159 & 0.0322 & 0.0540 & 0.0087 & 0.0127 & 0.0174  \\
    GAT    & 0.0283    & 0.0489    & 0.0636    & 0.0189    & 0.0240    & 0.0271  & 0.0233 & 0.0351 & 0.0463  & 0.0141 & 0.0170 & 0.0194  \\
    NGCF  & 0.0594    & 0.0999    & 0.1267    & 0.0364   & 0.0465    & 0.0522 & 0.0510 & 0.0732 & 0.0927  & 0.0321 & 0.0377 & 0.0418  \\
    LightGCN       & 0.0703    & 0.1105    & 0.1442    & 0.0405   & 0.0506    & 0.0578   & 0.0458 & 0.0697 & 0.0864 & 0.0296 & 0.0355 & 0.0391 \\
    \midrule
    Jodie      & 0.0270    & 0.0516    & 0.0732     & 0.0118    & 0.0179    & 0.0225    & 0.0253 & 0.0528 & 0.0783 & 0.0107 & 0.0176 & 0.0230 \\
    DyRep   & 0.0450    & 0.0826    & 0.1131    & 0.0213    & 0.0307    & 0.0372 & 0.0455 & 0.0899 & 0.1242 & 0.0199 & 0.0310 & 0.0383   \\
    TGN     & 0.0726    & 0.1276    & 0.1766    & 0.0338    & 0.0476    & 0.0580 & 0.0516 & 0.0923 & 0.1263 & 0.0248 & 0.0350 & 0.0422  \\
    DynShare   & 0.0750    & 0.1279    & 0.1718    & 0.0358    & 0.0491    & 0.0584  & 0.0563 & 0.1039 & 0.1440 & 0.0252 & 0.0371 & 0.0457  \\
    \midrule
    DynLLM    & \textbf{0.1052} & \textbf{0.1765} & \textbf{0.2321} & \textbf{0.0501}    & \textbf{0.0680} & \textbf{0.0799} & \textbf{0.0685} & \textbf{0.1120} & \textbf{0.1459} & \textbf{0.0340} & \textbf{0.0449} & \textbf{0.0521} \\
    \bottomrule
    \end{tabular}}
\end{table*}

\subsubsection{Baselines} We compare the DynLLM framework with a variety of competitive baselines, which can be categorized into two groups: static graph recommendation methods and dynamic graph recommendation methods. The details of these baselines are listed as follows:
\begin{itemize}
\item \textbf{GCN} \cite{kipf2016semi}: This is a classical static graph method that first introduces a localized first-order optimization of spectral graph convolutions.
\item \textbf{GAT} \cite{velivckovic2017graph}: It first employs attention mechanisms to weigh the importance of different neighbors in graph deep learning methods.
\item \textbf{NGCF} \cite{wang2019neural}: This method captures high-hop user-item interaction information and injects the CF signal into the graph embedding process. 
\item \textbf{LightGCN} \cite{he2020lightgcn}: It is a state-of-the-art GNN-based recommendation method that simplifies the design of GCN to make it more appropriate for recommendation task, including only neighborhood aggregation part for CF.
\item \textbf{JODIE} \cite{kumar2019predicting}: This approach utilizes a coupled RNN model to learn the embedding trajectory of users and items, and also introduces a novel linear projection operator to estimate the embedding of users in the future.
\item \textbf{DyRep} \cite{trivedi2019dyrep}: This work proposes a two-time scale deep temporal point process model to capture the continuous-time fine-grained temporal dynamics of association and communication processes.
\item \textbf{TGN} \cite{tgn_icml_grl2020}: This model exploits a novel combination of memory modules and temporal graph attention networks to evolve the dynamics in CTDGs, achieving state-of-the-art performance.
\item \textbf{DynShare} \cite{zhao2023time}: A state-of-the-art continuous time dynamic graph model that introduces a memory-based temporal graph attention networks to capture temporal structure information and leverages an innovative nonlinear projector to estimate user embeddings for next-time recommendation.
\end{itemize}

\begin{table}[htbp]
\centering
\caption{Ablation study on LLM-augmentation and distillation.}
\label{ablation llm}
\begin{tabular}{c|c|cccc}
\toprule
Dataset & Variants   &  w/o LLM  & w/o Distill & DynLLM  \\ 
\midrule
\multirow{8}{*}{Tmall} 
 & Recall@10             &  0.0342  &  0.0632   &  \textbf{0.1052} \\
& Recall@20             &  0.0697  &  0.1181   &  \textbf{0.1765} \\  
& Recall@30             &  0.1027  &  0.1660   &  \textbf{0.2321} \\
& NDCG@10            &  0.0148  &  0.0294  &  \textbf{0.0501} \\
& NDCG@20            &  0.0237  &  0.0431   &  \textbf{0.0680} \\
& NDCG@30            &  0.0307  &  0.0533   &  \textbf{0.0799} \\
\midrule
\multirow{8}{*}{Alibaba} 
& Recall@10             &  0.0178  &  0.0496   &  \textbf{0.0685} \\
& Recall@20             &  0.0365  &  0.0936   &  \textbf{0.1120} \\  
& Recall@30             &  0.0565  &  0.1331   &  \textbf{0.1459} \\
& NDCG@10            &  0.0080  &  0.0219   &  \textbf{0.0340} \\
& NDCG@20            &  0.0127  &  0.0330   &  \textbf{0.0449} \\
& NDCG@30            &  0.0169  &  0.0413   &  \textbf{0.0521} \\
\bottomrule
\end{tabular}
\end{table}

\subsubsection{Reproductive Settings} 
We implement the DynLLM framework using PyTorch\footnote{https://pytorch.org/} and run it with a single NVIDIA Tesla V100 GPU. For the LLMs, the LLM-generation model is selected as the "qwen-max-1201" chat completion model, and the LLM-embedding model uses the "text-embedding-v2" model in Alibaba Cloud\footnote{https://www.alibabacloud.com/}, with default settings as specified on the website. To ensure a fair comparison, we fix the dimension of user and item embeddings at 128 for DynLLM and all comparable baselines, where the item embedding is composed of 64-dimension static embedding and 64-dimension dynamic embedding. We employ a grid search strategy to optimize hyper-parameters and select the Adam optimizer \cite{kingma2014adam} for model parameter optimization, with a learning rate chosen from [1e-3, 5e-4, 1e-4] and the regularization coefficient $\lambda$ set to 1e-5. 
The batch size is set as 2048 and 4096 for Tmall and Alibaba dataset, respectively. For the TGAN component, we define the number of temporal graph attention layers as 2 and the number of heads as 2, while exploring the values for the number of temporal user neighbors $N_u$ and item neighbors $N_i$ from the set $\{2, 3, 4, 5\}$. Additionally, the distillation coefficient $r$ for the multi-faceted distilled attention mechanism is selected from $\{64, 80, 96, 112, 128\}$. Finally, we refer to the original papers and carefully tune the hyper-parameters of the baselines to suit our dataset.

\begin{figure*}[htbp]
\centering
\caption{Ablation study on multi-faceted profiles}
    \centering
    \subfloat{
    \includegraphics[width=0.245\textwidth]{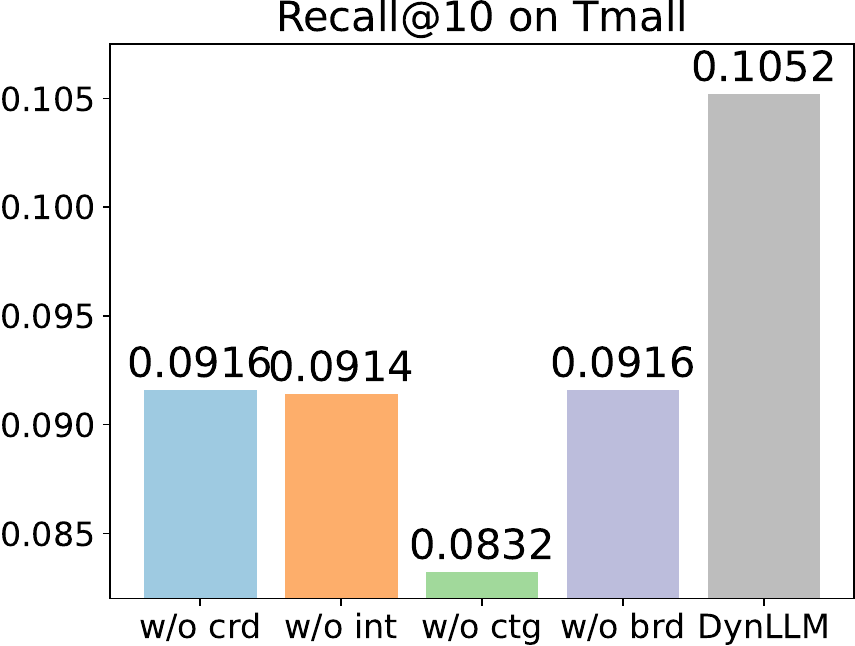}
    \label{ablation/Recall@10}
    }
    \subfloat{
    \includegraphics[width=0.245\textwidth]{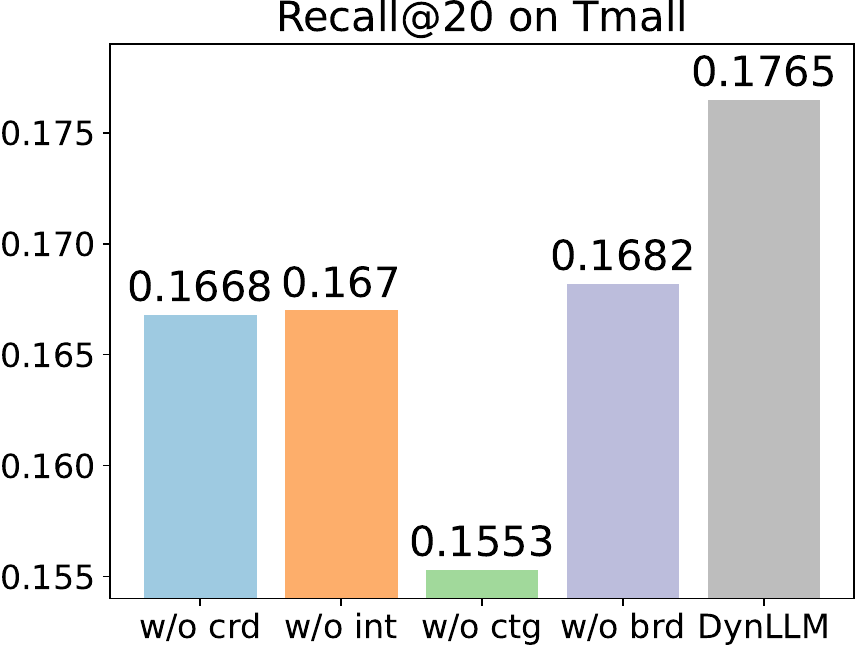}
    \label{ablation/Recall@20}
    }
    \subfloat{
    \includegraphics[width=0.245\textwidth]{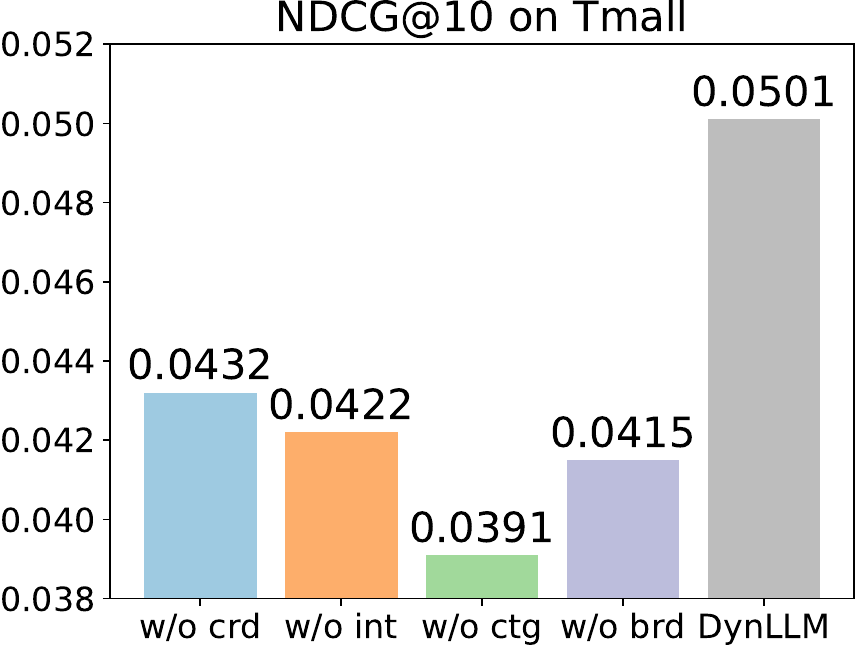}
    \label{ablation/Recall@30}
    }
    \subfloat{
    \includegraphics[width=0.245\textwidth]{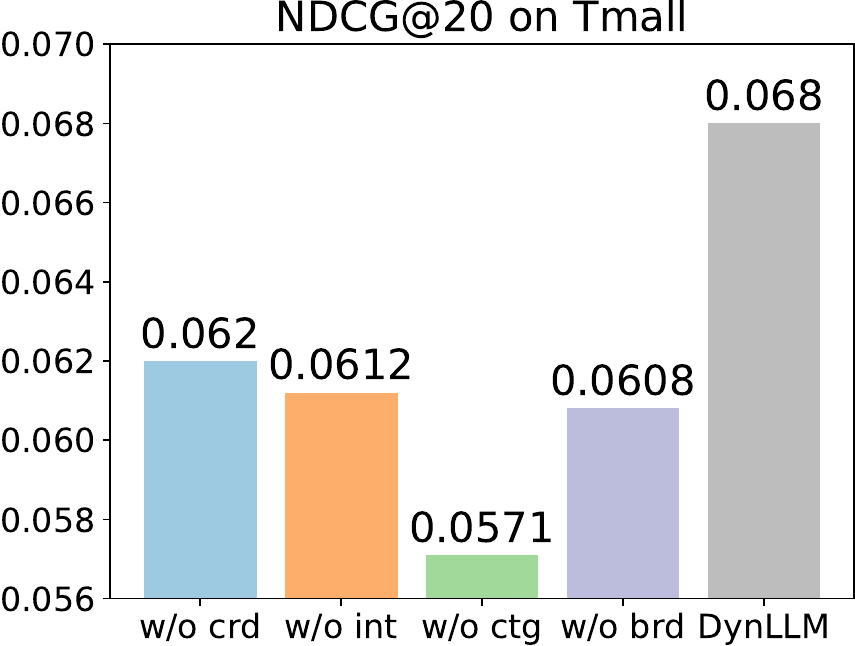}
    \label{ablation/Recall@50}
    } 
    \\
    \subfloat{
    \includegraphics[width=0.245\textwidth]{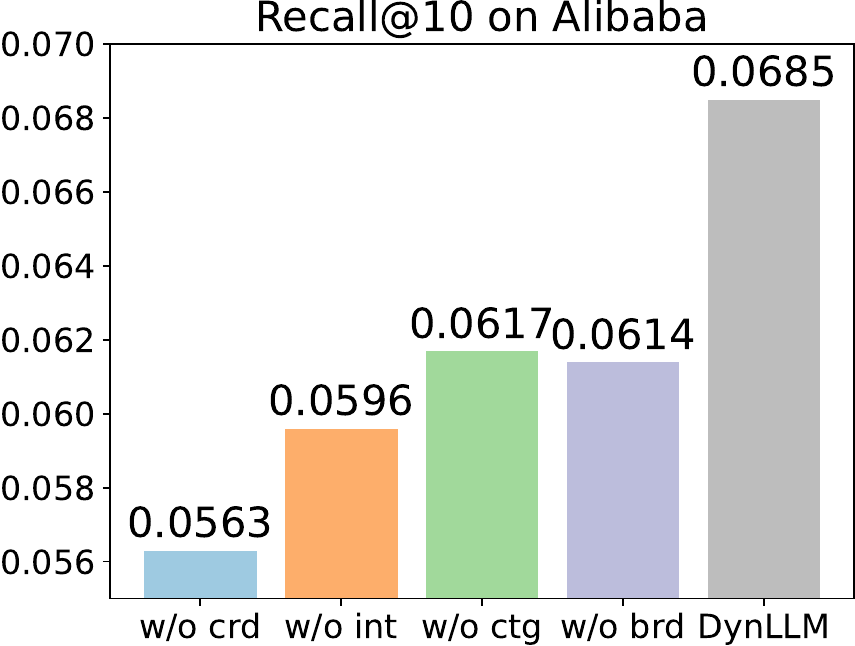}
    \label{ablation/NDCG@10}
    }
    \subfloat{
    \includegraphics[width=0.245\textwidth]{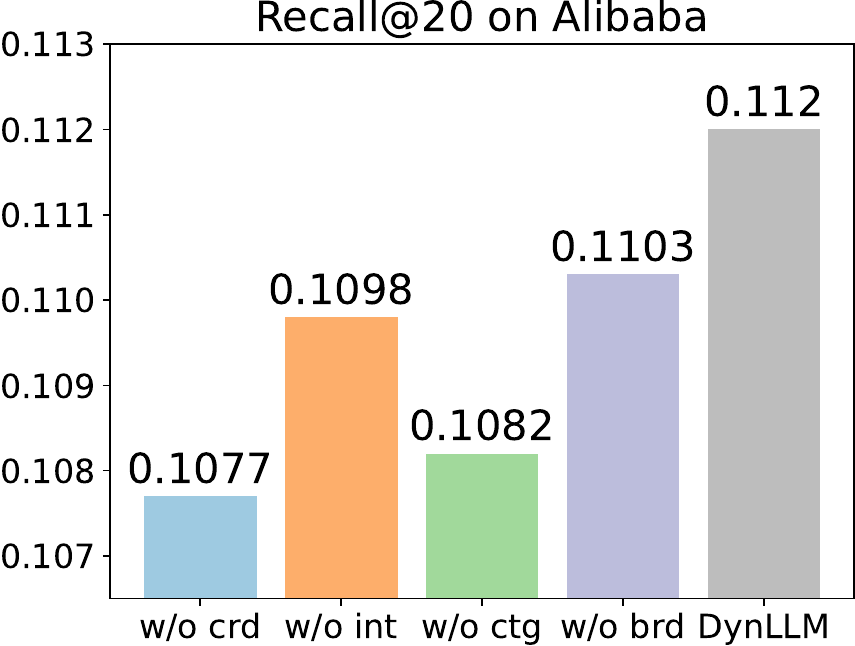}
    \label{ablation/NDCG@20}
    }
    \subfloat{
    \includegraphics[width=0.245\textwidth]{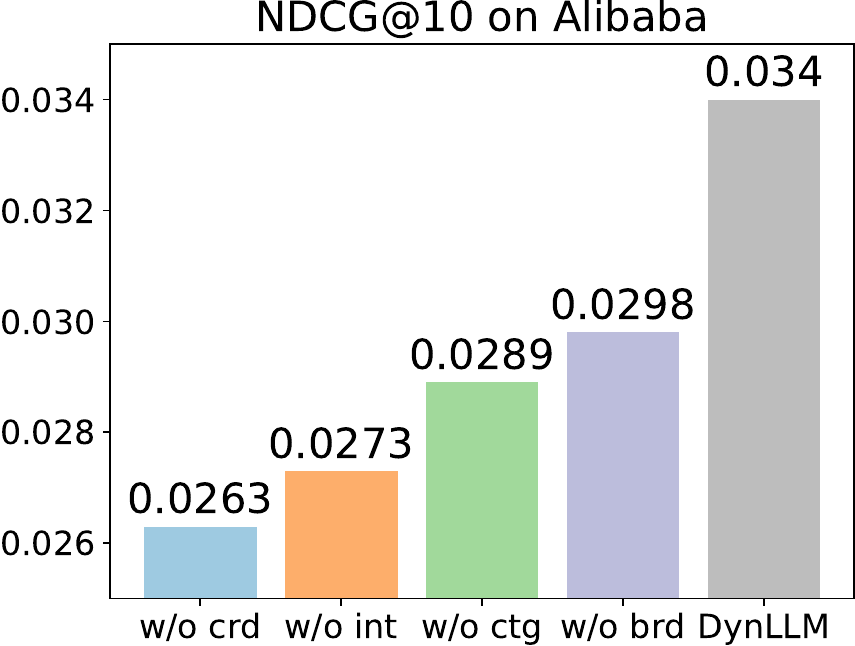}
    \label{ablation/NDCG@30}
    }
    \subfloat{
    \includegraphics[width=0.245\textwidth]{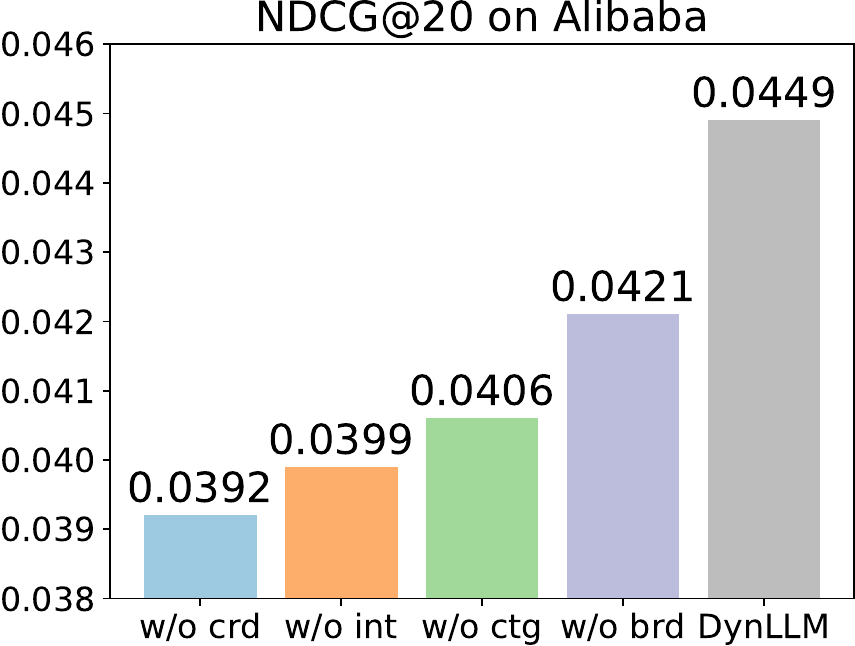}
    \label{ablation/NDCG@50}
    }
\label{ablation profile}
\end{figure*}

\subsubsection{Evaluation Protocols}
We evaluate our DynLLM approach in the Top-K recommendation task with two widely recognized evaluation metrics, i.e. NDCG (Normalized Discounted Cumulative Gain) and Recall. Specifically, we report Recall@K and NDCG@K of our method as well as all baselines, where $K = 10, 20, 30$. To ensure a comprehensive evaluation, we partition the interaction data chronologically, allocating the first 70\% data to train, the next 15\% to validate, and the final 15\% to test. And we adopt to rank the positive items against all items in validation and test sets. Afterward, we utilize the standard early-stopping strategy during the training stage by performing premature stopping if the Recall@10 metric on the validation dataset fails to improve for 10 consecutive epochs.

\subsection{Overall Performance (RQ 1)}
In this section, we conduct a comparative analysis of our DynLLM model against a diverse set of baseline methods in the context of Top-K recommendation performance, with K values set at 10, 20 and 30. The experimental results are presented in Table \ref{overall performance}, from which several key observations can be drawn:
\begin{itemize}
    \item Our DynLLM model consistently outperforms all existing state-of-the-art methods, and it can be broadly attributed to three aspects: (1) The explicit capture of fine-grained temporal information and graph topology information from both users and items through TGANs. (2) The enrichment and comprehensive revelation of implicit user information achieved through the LLM-augmented multi-faceted profiles, along with the introduction of a distilled attention mechanism that effectively integrates augmented information with temporal graph information. (3) The distillation of LLM-generated multi-faceted profile embeddings facilitates the preservation of relevant information while reducing the generated noise, thus improving the robustness of the DynLLM framework.
    \item In the majority of cases, dynamic graph recommendation methods exhibit better performance than their static counterparts. This highlights the pivotal role of temporal dynamics in accurately modeling user preferences for recommendation. However, there are notable exceptions, such as Jodie showing worse performance than NGCF and LightGCN, where Jodie only considers the temporal dynamics but neglects to capture high-order graph signals. On the contrary, TGN, and DynShare, which integrate fine-grained temporal information with high-order graph signals, surpass those static graph recommendation methods, except for Dyrep, which fails to utilize memory modules to store historical information.
    \item Among static graph recommendation methods, NGCF and LightGCN, which incorporate high-hop connectivity in user-item graphs into CF signals, yield superior results compared to simpler graph methods like GCN and GAT. These findings demonstrate the necessity and indispensability of exploiting the high-order connectivity into CF signals in recommendation tasks.
    \item Furthermore, DynShare outperforms other dynamic graph methods due to its memory module and novel nonlinear projector, enabling better prediction of future user embeddings and mitigating embedding staleness. On the other hand, our DynLLM results in superior performance over all dynamic methods, underscoring the effectiveness of LLMs in revealing implicit user profile information, ultimately leading to enhanced representation of dynamic user embedding.
\end{itemize}

\begin{figure*}[htbp]
\centering
\caption{Parameters sensitivity of the number of user neighbors and item neighbors}
    \centering
    \subfloat{
    \includegraphics[width=0.245\textwidth]{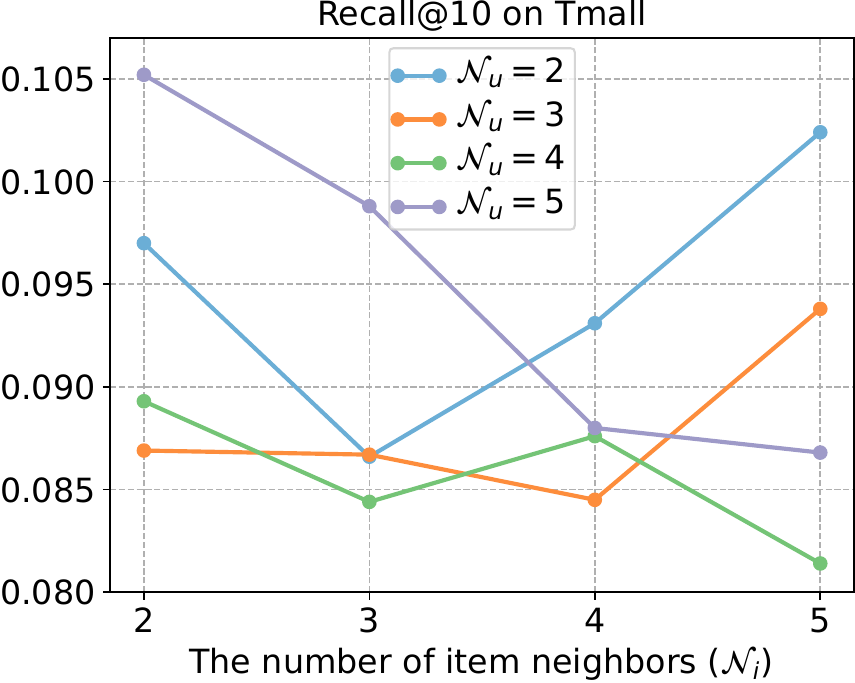}
    \label{parameters/Recall@10}
    }
    \subfloat{
    \includegraphics[width=0.245\textwidth]{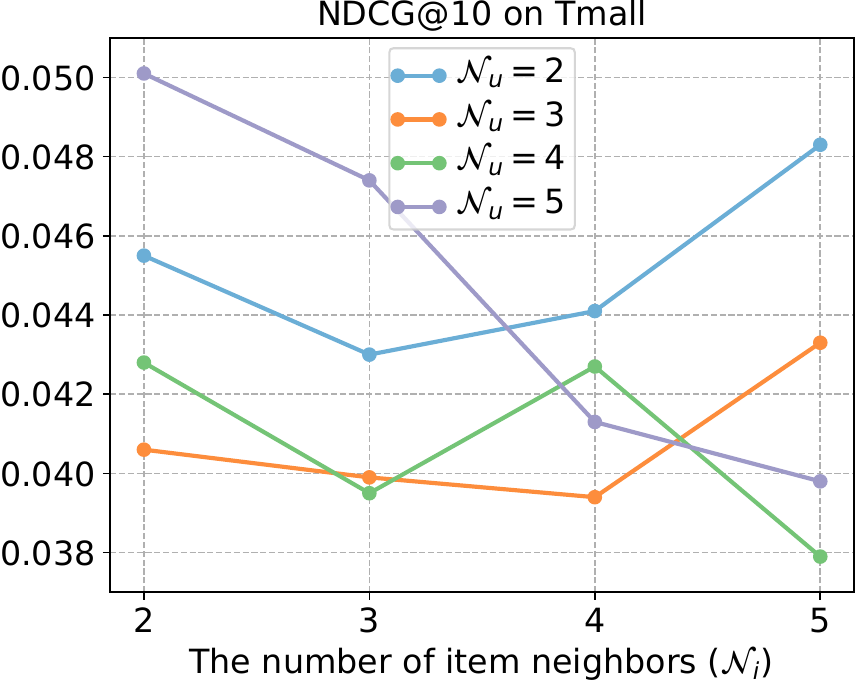}
    \label{parameters/NDCG@10}
    }
    \subfloat{
    \includegraphics[width=0.245\textwidth]{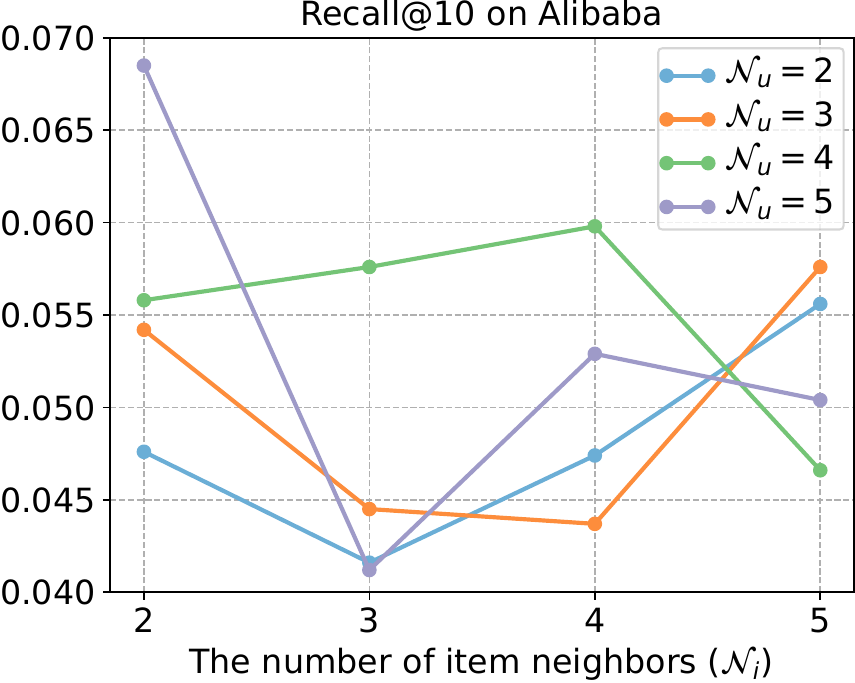}
    \label{parameters/Recall@10}
    }
    \subfloat{
    \includegraphics[width=0.245\textwidth]{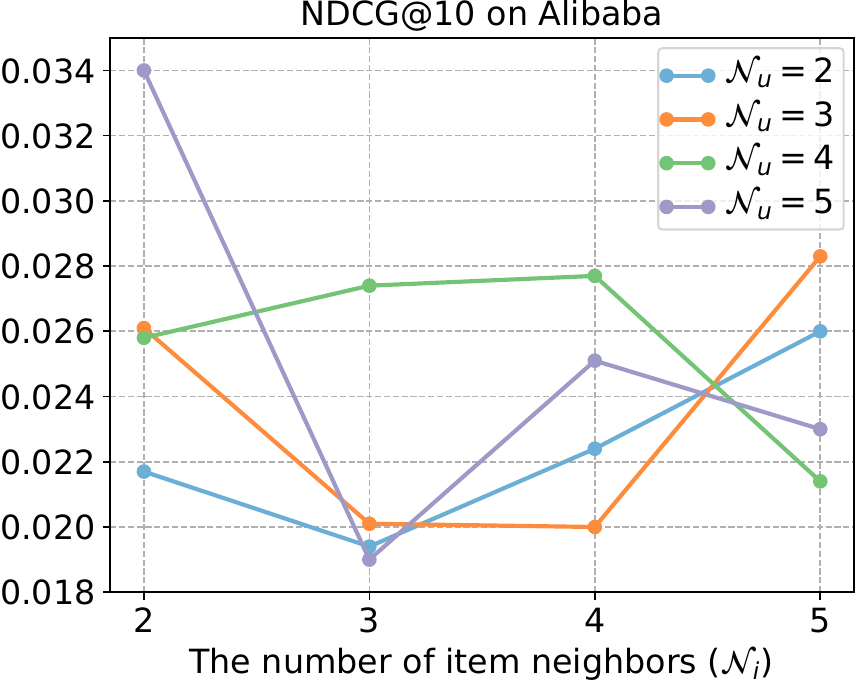}
    \label{parameters/NDCG@10}
    }
    
\label{parameter neighbor}
\end{figure*}

\begin{figure}[htbp]
\centering
\caption{Parameters sensitivity of the distillation coefficient}

    \centering
    \subfloat{
    \includegraphics[width=0.240\textwidth]{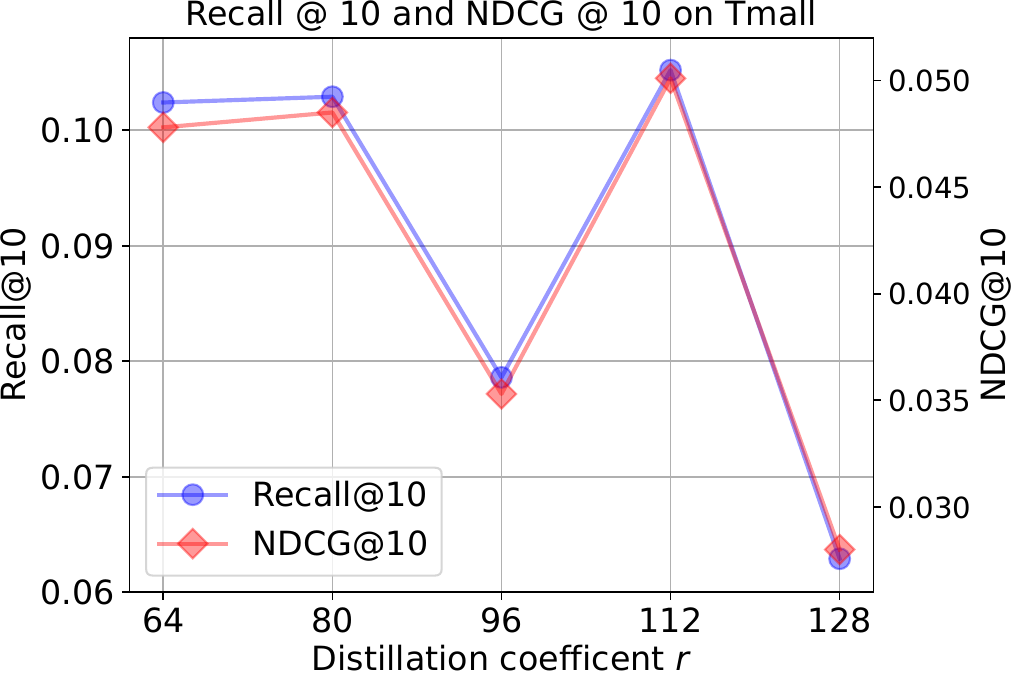}
    \label{parameters_prune/Tmall}
    }
    \subfloat{
    \includegraphics[width=0.240\textwidth]{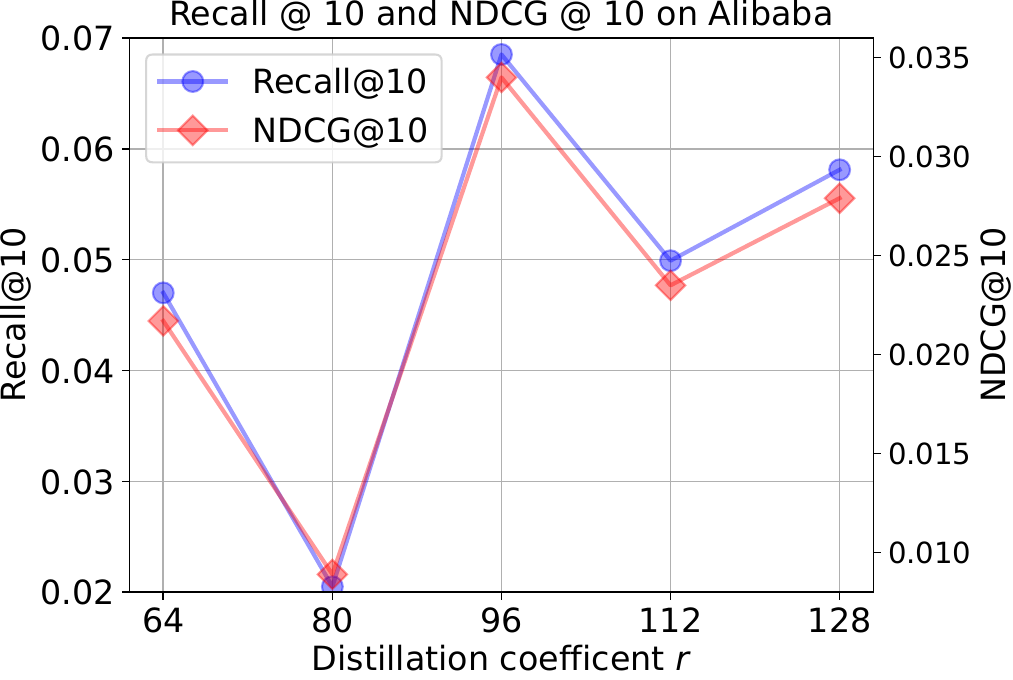}
    \label{parameters_prune/Alibaba}
    }
\label{parameter prune}
\end{figure}

\subsection{Ablation Study (RQ2 \& RQ3)}
In addition, to identify the specific contributions of each component of DynLLM, we conduct the following ablation study to demonstrate the impact of the proposed LLM-augmented profiles and multi-faceted distilled attention, and also to analyze how each facet of the profiles affects the overall performance.
\subsubsection{Effect of LLM-based Multi-facet Augmentation}
The LLM-based multi-faceted augmentation is a key component of our DynLLM model, which generates comprehensive profiles to enrich the underlying information of users. To emphasize the significance of the novel LLM-based multi-faceted augmentation, we create an ablation model that excludes LLMs augmentation and discards the augmented user profiles, denoted as \textit{w/o LLM}. Table \ref{ablation llm} reports the performance comparison between the original model and the ablation model without LLM. It is observed that the ablation model without LLMs exhibits a significant decrease in performance, underscoring the enriching role of LLM-augmented profiles in enhancing user profile semantics.

\subsubsection{Effect of Distilled Attention Mechanism}
To further investigate the necessity of the distilled attention mechanism, we design an ablation study \textit{w/o Distill} to verify its effectiveness. As depicted in Table \ref{ablation llm}, our DynLLM model outperforms the \textit{w/o Distill} variant, which indicates that the embedding generated directly from LLMs may contain noisy signals due to the bias of generated texts. Meanwhile, the novel design of distillation enhances the representation of LLM-generated embedding, emphasizing relevant signals while mitigating the impact of noisy signals. 

\subsubsection{Effect of Each Facet of Generated Profiles}
We now shift our focus to assess the effectiveness of each facet of LLM-generated multi-faceted profiles, by designing the following variants: \textit{w/o crd},  \textit{w/o int},  \textit{w/o ctg} and  \textit{w/o brd}, which neglect crowd segments, personal interests, preferred categories, and favored brands, respectively. From the experimental results shown in Figure \ref{ablation profile}, it is evident that our DynLLM surpasses all the variants, validating that each facet of LLM-generated multi-faceted profiles indeed enhances the expressiveness of user representation. It further shows that the variants \textit{w/o crd}, \textit{w/o int} and \textit{w/o ctg} yield the poorest performance in most cases, confirming that crowd segments, personal interests and preferred categories contribute more than favored brands. Thus, DynLLM provides an intuitive reflection of real-world dynamic recommendation scenarios.

\subsection{Hyper-parameter Sensitivity (RQ4 \& RQ5)}
Additionally, we conduct experiments to explore how the number of neighbors (i.e., $\mathcal{N}_u$ and  $\mathcal{N}_i$) and the distillation coefficient (i.e., $r$) affect our DynLLM performance. 
\subsubsection{Effect of Neighbors Number}
We first investigate the impact of different numbers of neighbors $\mathcal{N}_u$ and $\mathcal{N}_i$. Specifically, we tune $\mathcal{N}_u$ and $\mathcal{N}_i$ within the range of $\{2, 3, 4, 5\}$ while maintaining other hyper-parameters at their optimal settings. As illustrated in Figure \ref{parameter neighbor}, DynLLM approaches optimal performance when $\mathcal{N}_u = 5$ and $\mathcal{N}_i = 2$ for both datasets, which is similar to the ratio of the user's average interactions and item's average interactions. From this observation, we select these optimal values in the experiments to ensure the best performance of model.

\subsubsection{Effect of Distillation Coefficient}
Furthermore, DynLLM utilizes a distillation coefficient $r$ to mitigate noisy signals in LLM-generated embeddings. We explore a range of values, specifically $\{64, 80, 96, 112, 128\}$, to determine optimal performance on both datasets. Figure \ref{parameter prune} summarizes the overall results obtained from tuning the distillation coefficient. Notably, we consistently achieve optimal performance when $r$ is fixed at 112 and 96 on Tmall and Alibaba dataset, respectively. This observation suggests that assigning appropriate distillation coefficient can effectively minimize LLM-generated noise and enhance the performance.

\section{Conclusion}
In this paper, we introduced a novel task, LLM-augmented dynamic recommendation via continuous time dynamic graphs, and proposed the DynLLM model to effectively integrate LLM-augmented information with temporal graph information. Specifically, we first leveraged temporal graph attention networks to capture the fine-grained temporal dynamics and graph topology information from the perspective of users and items. Subsequently, utilizing Large Language Models (LLMs), we generated multi-faceted profiles to augment and enrich user information, extracting potential correlations between users and items. Afterward, we also developed an innovative multi-faceted distilled attention mechanism to enhance the expressiveness of LLM-generated user profiles, which facilitated the fusion of temporal graph embedding and refined profile embeddings. As a result, our approach not only successfully captured graph topology information along with fine-grained temporal information, but also gradually incorporated the LLM-augmented information to prepare for next-time recommendation. Finally, extensive experiments on two real-world online e-commerce datasets with rich texts have demonstrated the superiority of DynLLM compared to competitive baseline methods.

\bibliographystyle{ACM-Reference-Format}
\bibliography{dynllm}

\end{sloppypar}
\end{document}